\documentclass{elsart}

\usepackage[square,comma]{natbib}
\usepackage{graphicx}
\usepackage{pxfonts}
\usepackage{lineno}

\usepackage{amssymb}

\journal{}

\begin{document}

\thispagestyle{empty}
\begin{Large}
\textbf{DEUTSCHES ELEKTRONEN-SYNCHROTRON}

\textbf{\large{Ein Forschungszentrum der
Helmholtz-Gemeinschaft}\\}
\end{Large}

DESY 10-010

January 2010

\begin{eqnarray}
\nonumber &&\cr \nonumber && \cr \nonumber &&\cr
\end{eqnarray}
\begin{eqnarray}
\nonumber
\end{eqnarray}
\begin{center}
\begin{Large}
\textbf{Control of the amplification process in baseline XFEL
undulator with mechanical SASE switchers}
\end{Large}
\begin{eqnarray}
\nonumber &&\cr \nonumber && \cr
\end{eqnarray}

\begin{large}
Gianluca Geloni,
\end{large}
\textsl{\\European XFEL GmbH, Hamburg}
\begin{large}

Vitali Kocharyan and Evgeni Saldin
\end{large}
\textsl{\\Deutsches Elektronen-Synchrotron DESY, Hamburg}
\begin{eqnarray}
\nonumber
\end{eqnarray}
\begin{eqnarray}
\nonumber
\end{eqnarray}
ISSN 0418-9833
\begin{eqnarray}
\nonumber
\end{eqnarray}
\begin{large}
\textbf{NOTKESTRASSE 85 - 22607 HAMBURG}
\end{large}
\end{center}
\clearpage
\newpage

\begin{frontmatter}



\title{Control of the amplification process in baseline XFEL undulator with mechanical SASE switchers}


\author[XFEL]{Gianluca Geloni\thanksref{corr},}
\thanks[corr]{Corresponding Author. Tel: ++49 40 8998 5450. Fax: ++49 40 8998 1905. E-mail address: gianluca.geloni@xfel.eu}
\author[DESY]{Vitali Kocharyan}
\author[DESY]{and Evgeni Saldin}

\address[XFEL]{European XFEL GmbH, Hamburg, Germany}
\address[DESY]{Deutsches Elektronen-Synchrotron (DESY), Hamburg,
Germany}

\begin{abstract}
The magnetic gap of the baseline XFEL undulators can be varied
mechanically for wavelength tuning. In particular, the wavelength
range $0.1$ nm - $0.4$ nm can be covered by operating the European
XFEL with the SASE2 undulator. The length of the SASE2 undulator
($256.2$ m) is sufficient to independently generate three pulses
of different radiation wavelengths at saturation. Normally, if a
SASE FEL operates in saturation, the quality of the electron beam
is too bad for generation of SASE radiation in the subsequent part
of undulator which is resonant at a few times longer wavelength.
The new method of  SASE undulator-switching  based on the rapid
switching of the FEL amplification process proposed in this  paper
is an attempt to get around this obstacle.  Using mechanical SASE
shutters installed within short magnetic chicanes in the baseline
undulator, it is possible to rapidly switch the FEL photon beam
from one wavelength to another, providing simultaneous multi-color
capability. Combining this method with a photon-beam distribution
system can provide an efficient way to generate a multi-user
facility.
\end{abstract}

%
%

\end{frontmatter}



\section{\label{sec:intro} Introduction}

The recent achievement of LCLS \cite{LCLS1,LCLS2} relies on a
high-performance beam formation system, which works as in the
ideal operation scenario described in the conceptual design report
\cite{LCLS1}. In particular, the small electron-beam emittance
achieved allows saturation within $20$ undulator modules, out of
the $33$ available. A similar scenario is also foreseen for the
European XFEL. One has, then, the possibility of taking advantage
of a long, unused part of the SASE undulators to upgrade the
facility.

In this paper we describe a method for providing simultaneous
multi-color capability at three different wavelengths. For the
sake of exemplification we will consider radiation around $0.2$
nm, $0.15$ nm and $0.1$ nm, produced by an electron beam with
$0.4~\mu$m normalized emittance and $0.25$ nC charge. After
integration with a photon-beam distribution system, this method
can lead to an effcient multi-user facility.

\begin{figure}[tb]
\includegraphics[width=1.0\textwidth]{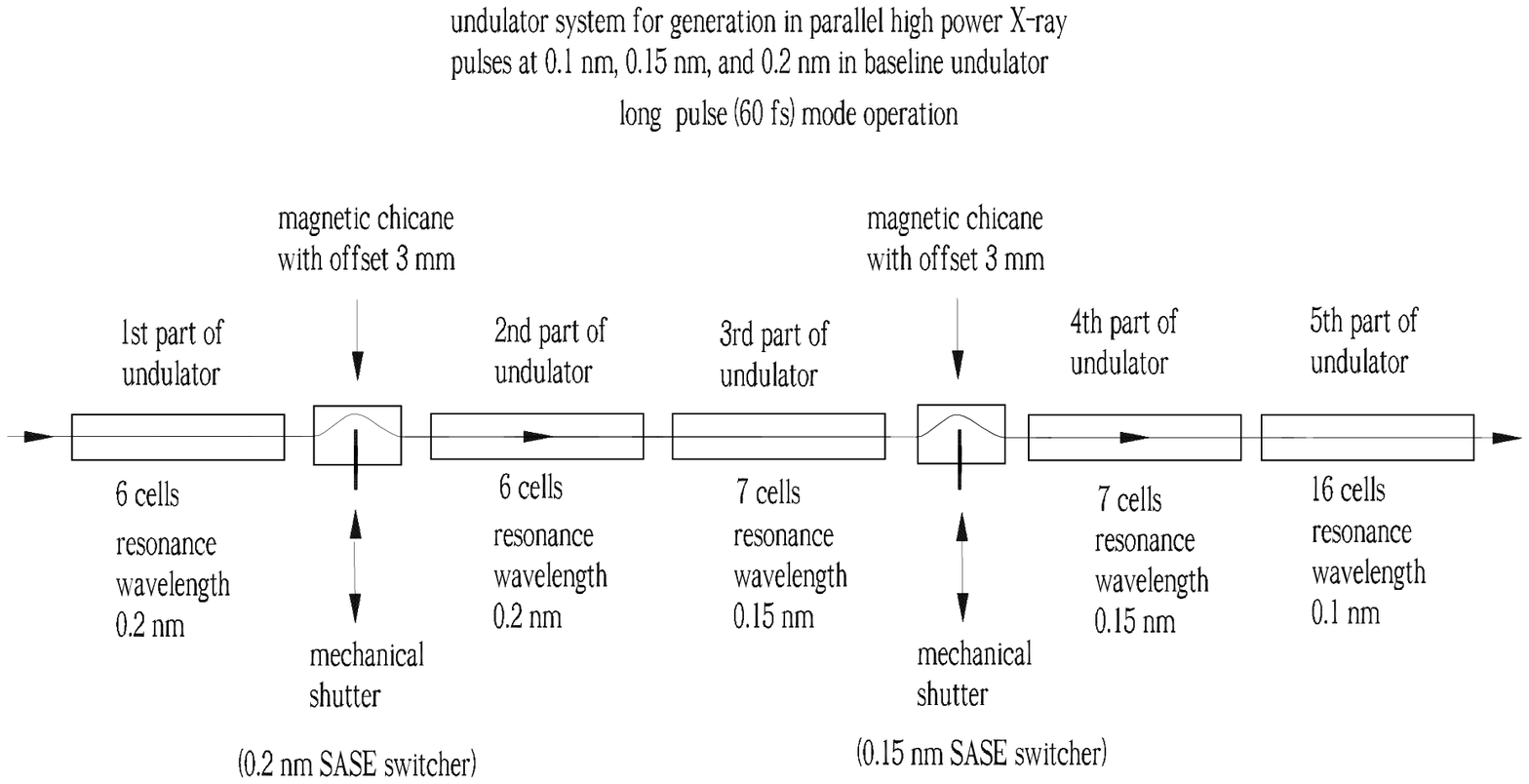}
\caption{Design of the undulator system for the three color X-ray
source in the case of a long-pulse ($60$ fs) mode operation.
Wavelength selection is based on the use of two mechanical SASE
switchers. Each SASE switcher consists of a short magnetic chicane
and a SASE shutter. The magnetic chicane generates an offset for
the SASE shutter installation and additionally washes out the
electron beam modulation. The extra path-length is much smaller
than the radiation pulse length.} \label{s1}
\end{figure}

\begin{figure}[tb]
\includegraphics[width=1.0\textwidth]{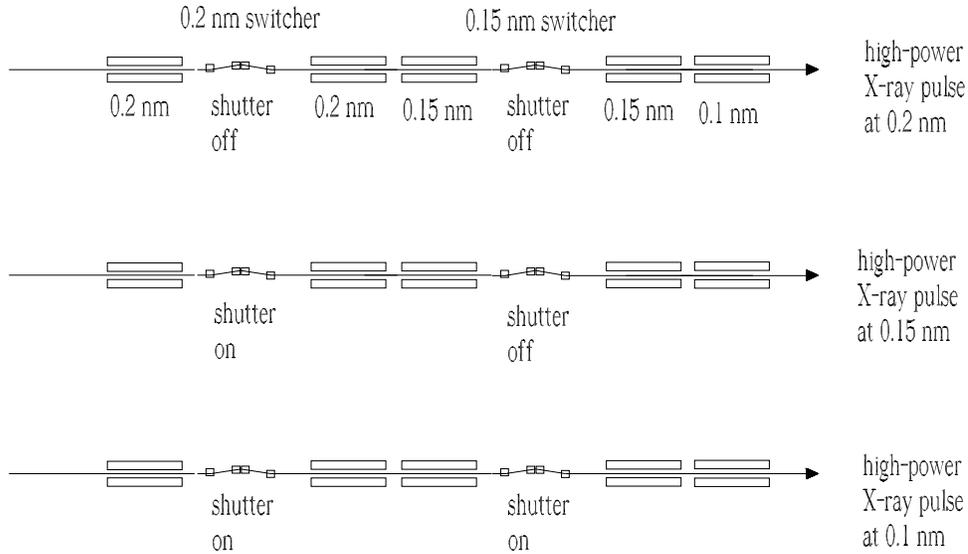}
\caption{The three modes of operation for the SASE shutters in the
baseline undulator.} \label{s2}
\end{figure}
\begin{figure}[tb]
\includegraphics[width=1.0\textwidth]{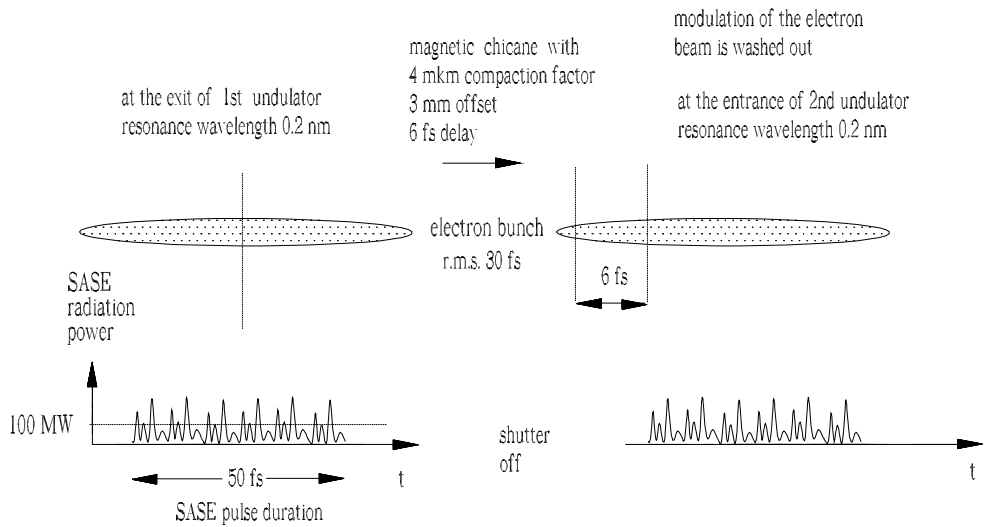}
\caption{Sketch of principle of multi-color X-ray pulse generation
in the baseline XFEL undulator. Here the SASE shutter is off.
Modulation of the electron beam due to the FEL process in the
first undulator part is washed out in the magnetic chicane. In the
second part of the undulator, the seeded main part of electron
bunch reaches saturation with ten GW power level at $0.2$ nm
wavelength.} \label{s3}
\end{figure}
\begin{figure}[tb]
\includegraphics[width=1.0\textwidth]{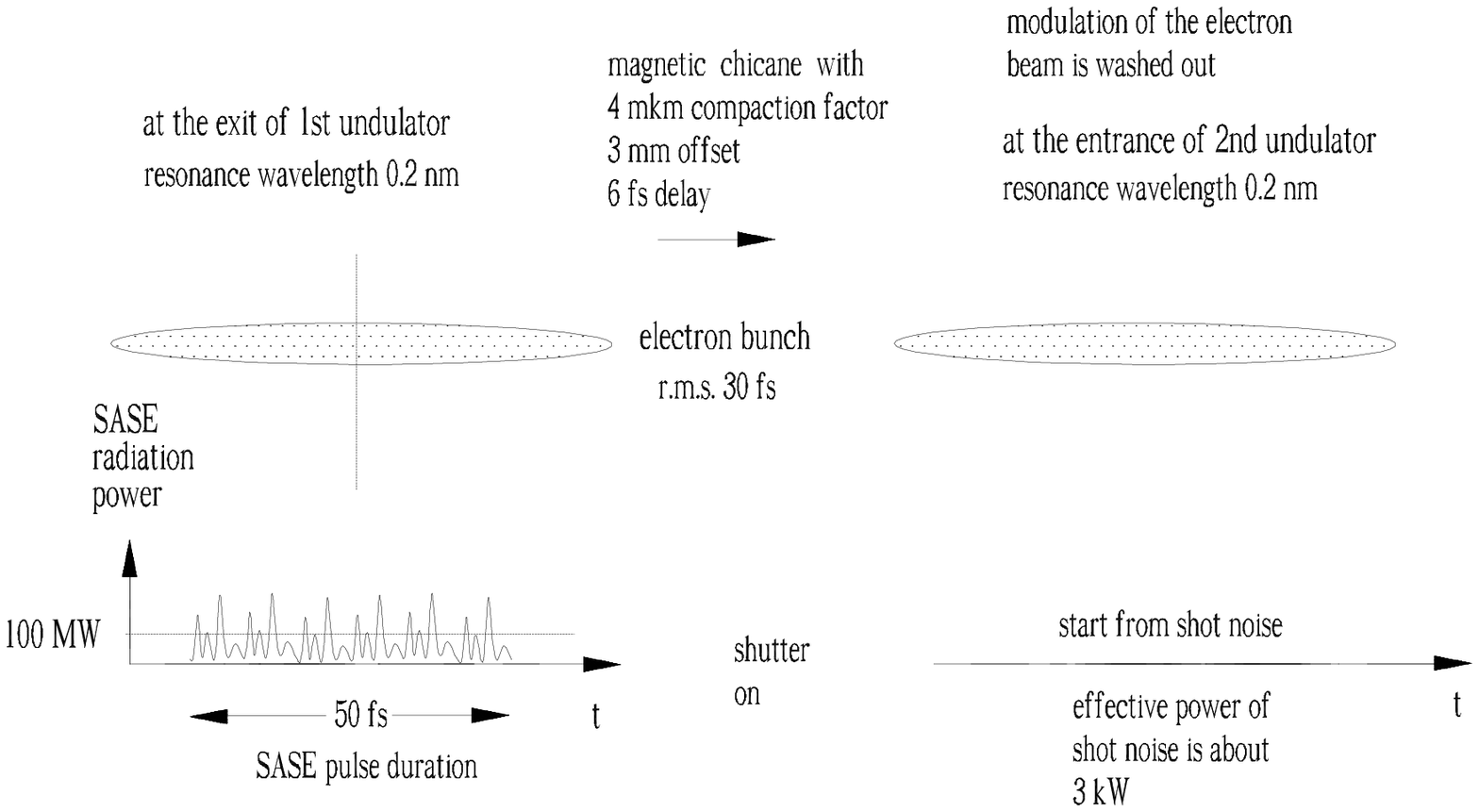}
\caption{Sketch of principle of multi-color X-ray pulse generation
in the baseline XFEL undulator. Here the SASE shutter is on.
Modulation of electron bunch due to FEL process in the first part
of undulator is washed out in magnetic chicane. In the second part
of undulator FEL amplification process start up from shot noise
and reaches 0.1 GW power level only. As a result, energy losses
and energy spread within the electron bunch are negligible and the
electron bunch is still a good "active medium" for the next color
pulse generation in the following undulator parts.} \label{s4}
\end{figure}
In its essence, the method is based on controlling the
amplification process in the baseline XFEL undulator with the help
of SASE switchers constituted by mechanical shutters to be
installed at the position of a weak magnetic chicanes at specific
locations down the SASE undulator. A sketch of the setup is
presented in Fig. \ref{s1}. Three different modes of operations
are foreseen, based on a long ($60$ fs) pulse, depending on what
shutters are off and on (see Fig. \ref{s2}). In the first part of
the undulator the beam undergoes the SASE process in the linear
regime. After that, it passes through the first switcher, which
consists of two devices. First, a weak magnetic chicane creating a
transverse offset for the electrons and washing out the electron
beam modulation. Second, a mechanical shutter, which has two
positions: "on" when absorbing SASE radiation and "off", when SASE
radiation from the first undulator passes unperturbed to the
second undulator.  The electron bunch passes through the magnetic
chicane, and the beam modulation is washed out. Note that in the
chicane, the electron bunch is also delayed with respect to the
radiation but the extra-path length can be chosen small enough to
provide a shift of $6$ fs, much shorter than the long electron
bunch ($30$ fs rms). If the shutter is off (see Fig. \ref{s2} top
and Fig. \ref{s3}),  at the position of the second undulator
almost all the "fresh" (washed out) bunch is seeded by the
radiation coming from the first part, and the seeded main part of
electron bunch reaches saturation with ten GW power level at $0.2$
nm wavelength. In this case, energy losses and energy spread
within the electron bunch are important, and the bunch cannot
further be used for generation of high intensity SASE radiation at
wavelength  comparable to $0.1$ nm. If the shutter is on (see Fig.
\ref{s2} middle and bottom, and Fig. \ref{s4}) instead, a fresh
electron beam will enter the second part of the undulator, and
will start radiate according to the usual SASE process. The second
part of the undulator is short enough so that the SASE
amplification process ends, once more, in the linear regime. In
this case energy losses and energy spread within the electron
bunch are negligible and the electron bunch is still a good
"active medium" for the next color pulse generation in the
following undulator parts. These parts are tuned to a different
wavelength, namely $0.15$ nm and $0.1$ nm. When the second shutter
is off (see Fig. \ref{s2} middle), the seeded SASE process reaches
saturation at $0.15$ nm. When the second shutter is on, instead,
one has saturation in the final part of the undulator only at
$0.1$ nm (see Fig. \ref{s2} bottom). The distribution of photons
can be achieved on the basis of pulse trains, thereby allowing
many users working in parallel at different wavelengths. For the
temporal structure of the radiation produced at the European XFEL
\cite{tdr-2006}, this means that we have no restrictive
requirement on the switching time of the mechanical shutter, which
should be simply shorter than $100$ msec, a very suitable
condition for mechanical system.

\begin{table}
\caption{Parameters for the pulse mode used in this paper. The
undulator parameters are the same of those for the European XFEL,
SASE2, at 17.5 GeV electron energy.}

\begin{small}\begin{tabular}{ l c c}
\hline
& ~ Units &  ~ \\
\hline
Undulator period      & mm                  & 47.9   \\
Undulator length      & m                   & 256.2  \\
Undulator segment length        & m                   & 5.0    \\
Intersection length             & m                   & 1.1    \\
Number of segments    & -                   & 42     \\
K parameter (rms)     & -                   & 1.97-2.96  \\
$\beta$               & m                   & 17     \\
Wavelength            & nm                  & 0.1 - 0.2   \\
Energy                & GeV                 & 17.5   \\
Charge                & nC                  & 0.25  \\
Bunch length (rms)    & $\mu$m              & 10.0    \\
Normalized emittance  & mm~mrad             & 0.4    \\
Energy spread         & MeV                 & 1.5    \\
\hline
\end{tabular}\end{small}
\label{tab:fel-par}
\end{table}
In the next section we present a feasibility study of the method,
and we make exemplifications with the parameters of the SASE2 line
of the European XFEL (see Table \ref{tab:fel-par}). With the help
of this scheme it will be possible to provide in parallel X-rays
at three different wavelengths around 0.2 nm, 0.15 nm and 0.1 nm.
In the following section we will see how, combining this method
with a photon-beam distribution system, one can provide an
efficient way to generate a multi-user facility.

\section{Feasibility study}

In the following we describe the outcomes of computer simulations
using the code Genesis $1.3$ \cite{GENE}.

\subsection{Both shutters off}

First we consider the case when both shutters are off and the
first and second parts of the undulator are tuned at $0.2$ nm,
Fig. \ref{s2} top. 

\begin{figure}[tb]
\includegraphics[width=1.0\textwidth]{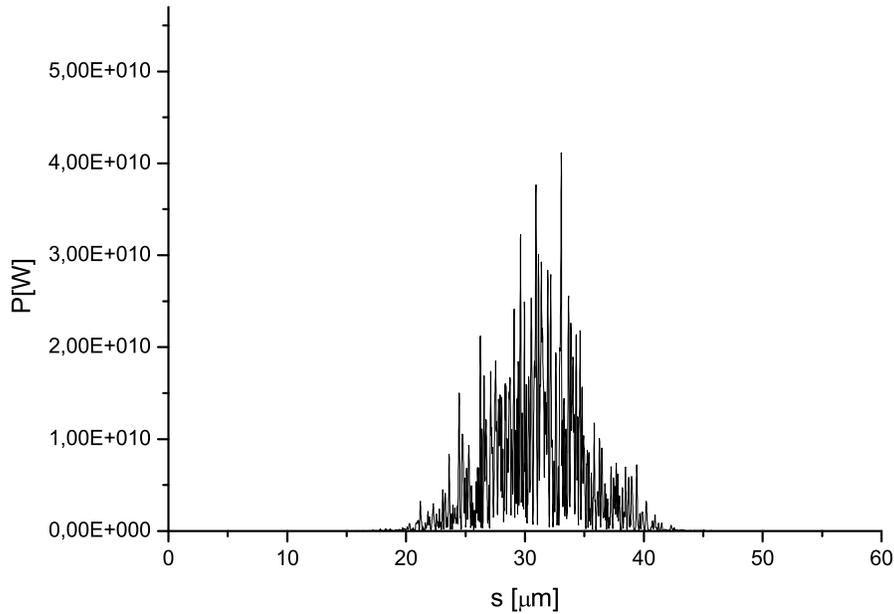}
\caption{Beam power distribution at the end of the second part of
the undulator after $7+7$ cells ($42.7$ m+$42.7$ m). The first
shutter is off.} \label{1pow}
\end{figure}

\begin{figure}[tb]
\includegraphics[width=1.0\textwidth]{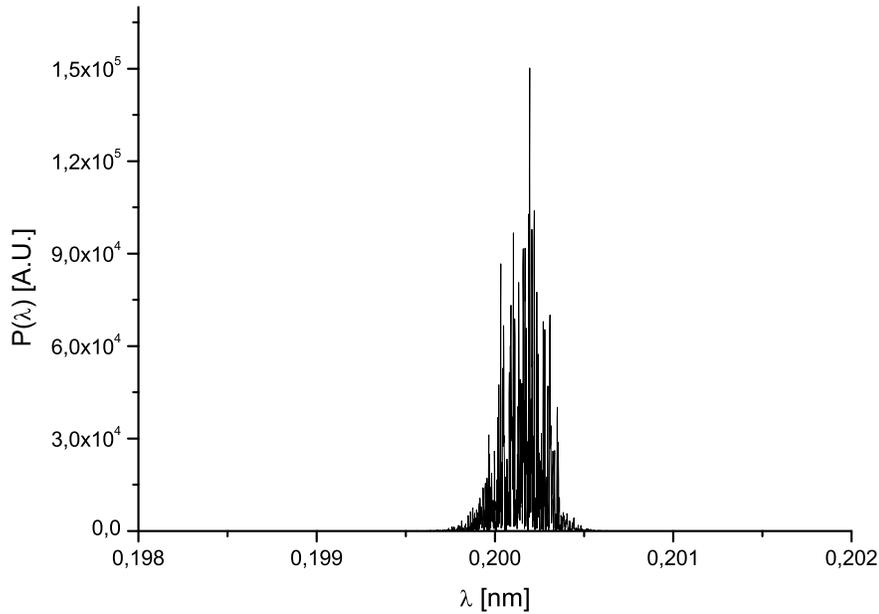}
\caption{Spectrum at the end of the second part of the undulator
$7+7$ cells ($42.7$ m+$42.7$ m). The first shutter is off.}
\label{1spe}
\end{figure}

The power and spectrum after the second part of
the undulator are shown in Fig. \ref{1pow} and \ref{1spe}.

\begin{figure}[tb]
\includegraphics[width=0.5\textwidth]{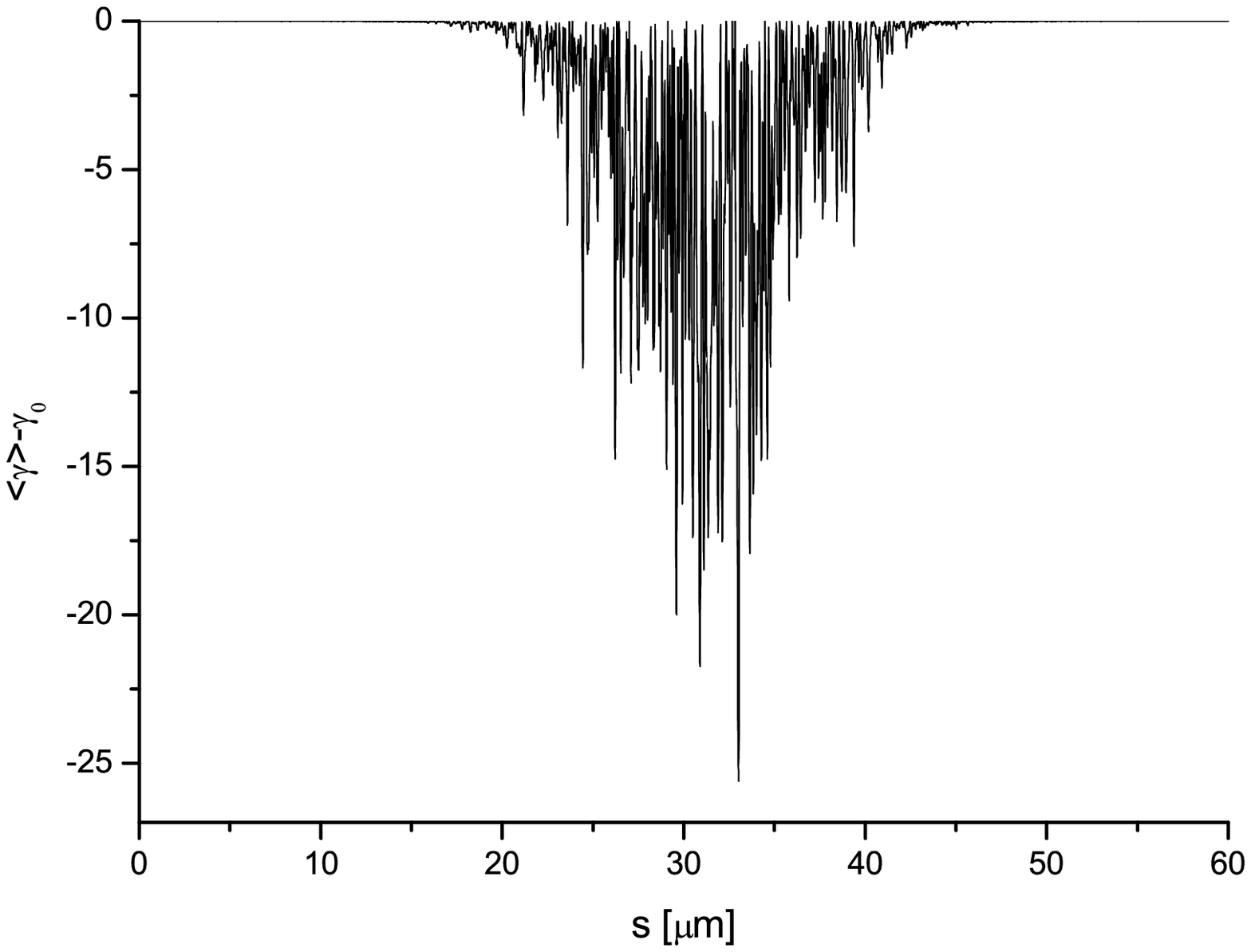}
\includegraphics[width=0.5\textwidth]{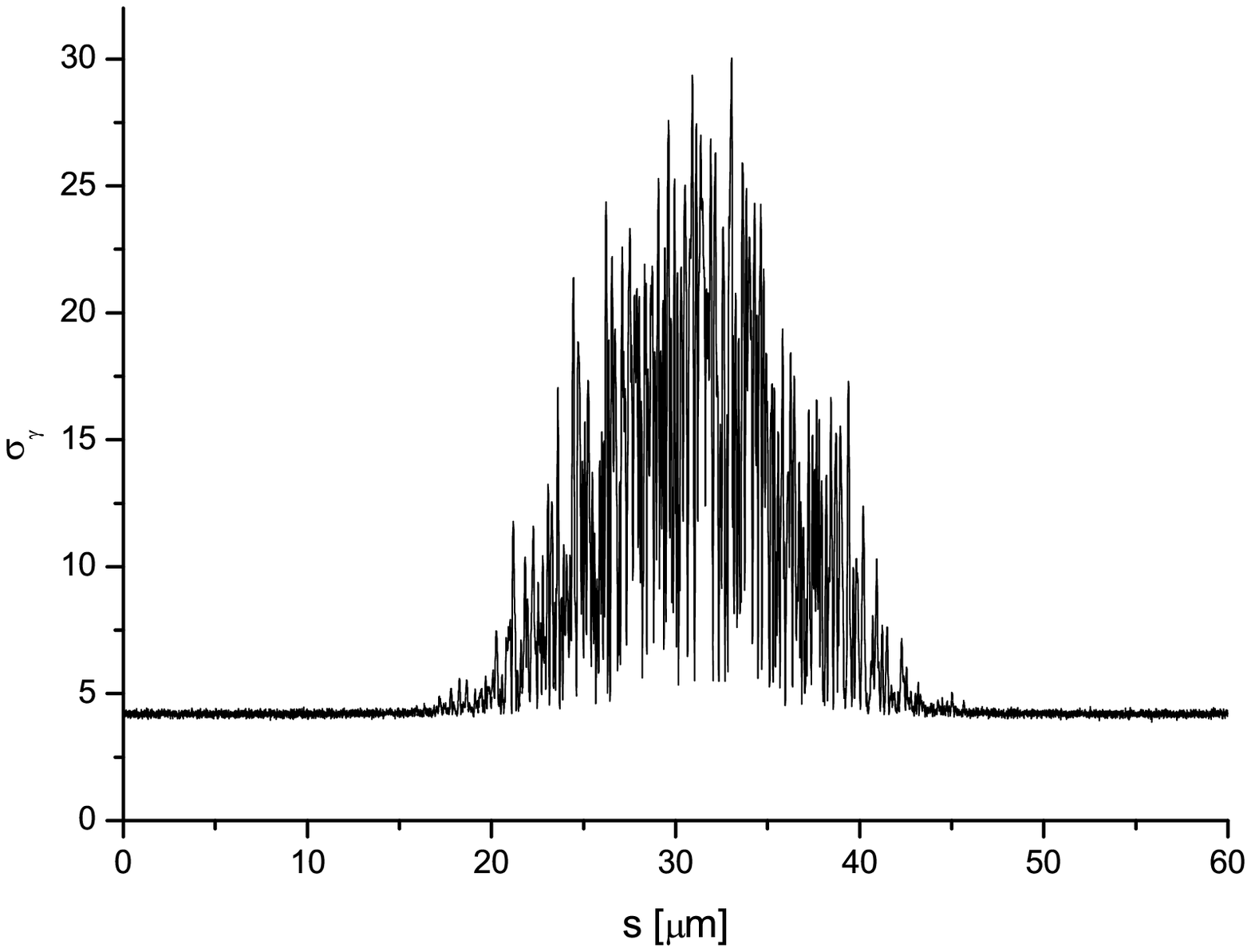}
\caption{Electron beam energy loss (left) and induced energy
spread (right) at the entrance of the third part of the undulator.
The first shutter is off.} \label{1enspr}
\end{figure}
In this case the SASE process reaches saturation in the second
undulator part. The electron beam energy loss and induced energy
spread are severe, and prevent the beam to undergo the SASE
process again in the following undulator parts, Fig. \ref{1enspr}.

\subsection{First shutter on}

\begin{figure}[tb]
\includegraphics[width=0.5\textwidth]{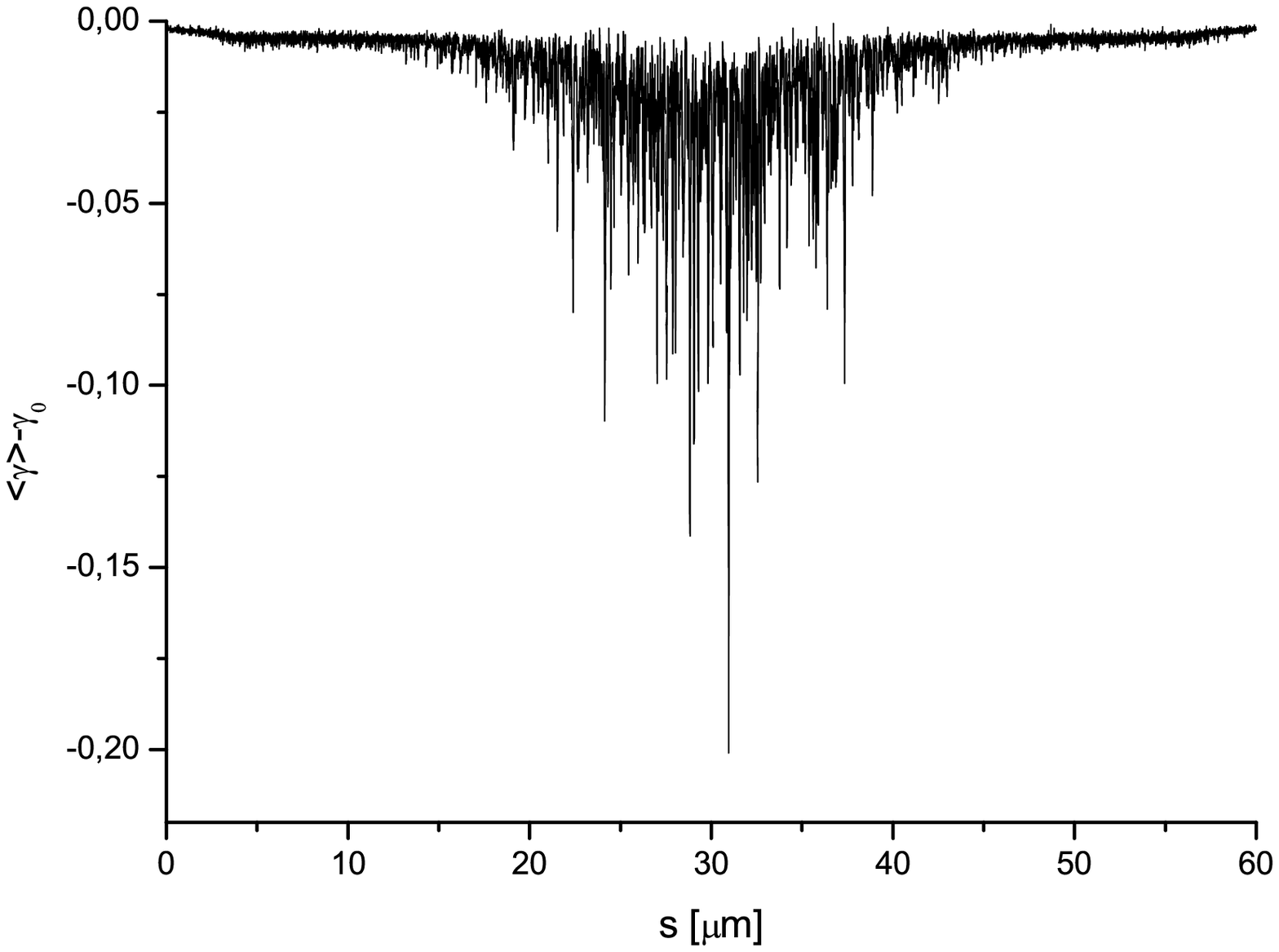}
\includegraphics[width=0.5\textwidth]{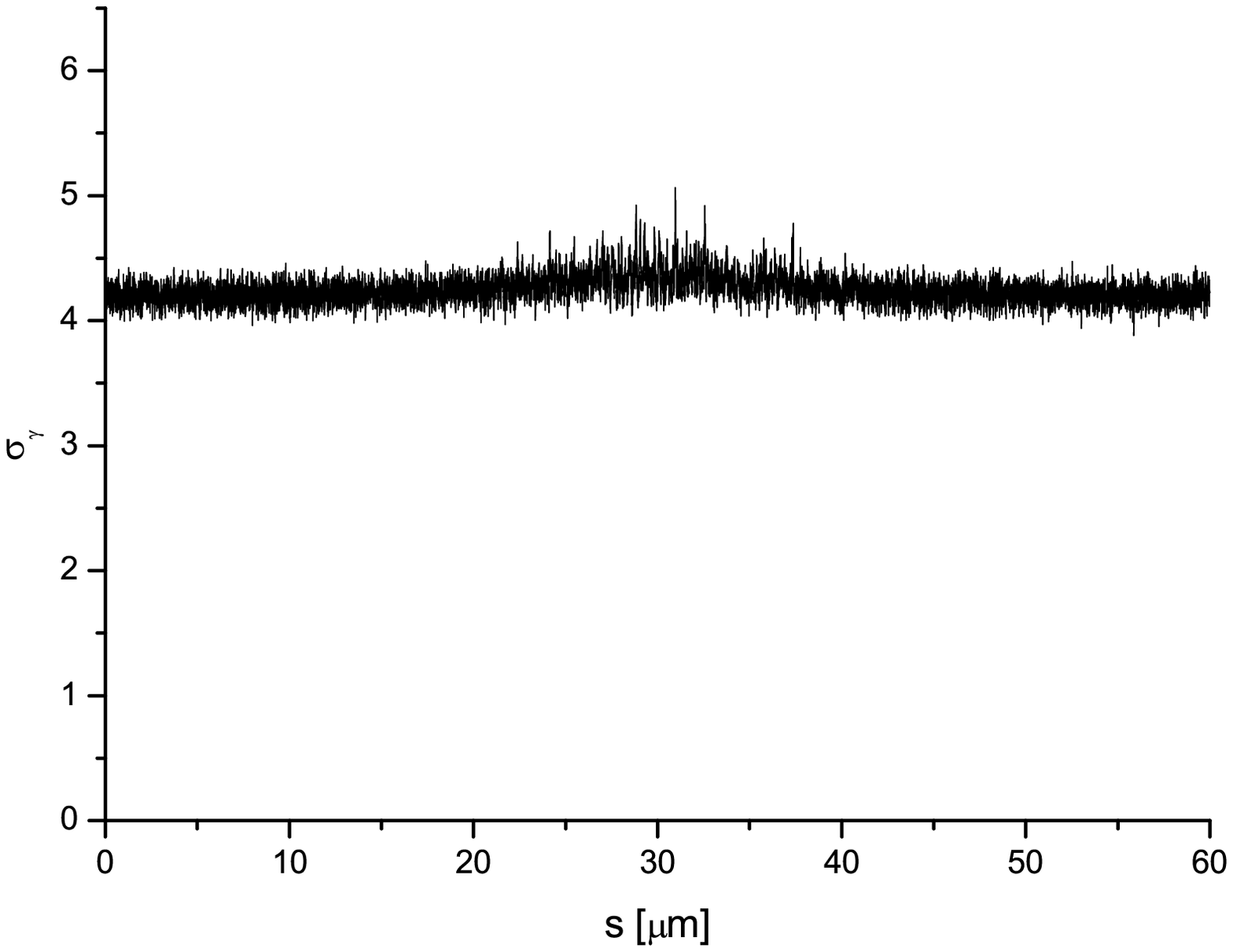}
\caption{Electron beam energy loss (left) and induced energy
spread (right) at the entrance of the third part of the undulator.
The first shutter is on.} \label{2enspr}
\end{figure}

Subsequently, we studied the case when the first shutter is on and
the second is off, corresponding to the situation in Fig. \ref{s2}
middle. The presence of the shutter prevents the seeding process
in the second undulator part. As a result the SASE process at
$0.2$ nm is far from saturation, and the beam can be used to
produce radiation at $0.15$ nm in the third (6 segments, $36.6$ m)
and fourth part (6 segments, $36.6$ m) of the undulator. In this
case, the properties of the electron beam at the entrance of third
part of the undulator are summarized in Fig. \ref{2enspr}.

%

\begin{figure}[tb]
\includegraphics[width=1.0\textwidth]{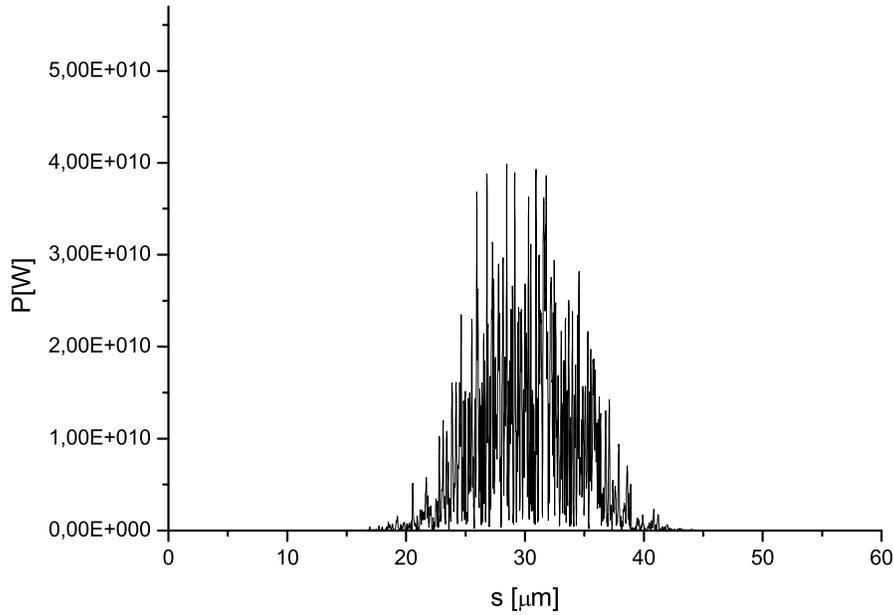}
\caption{Beam power distribution at the end of the fourth part of
the undulator. The first shutter is on.} \label{2pow}
\end{figure}

\begin{figure}[tb]
\includegraphics[width=1.0\textwidth]{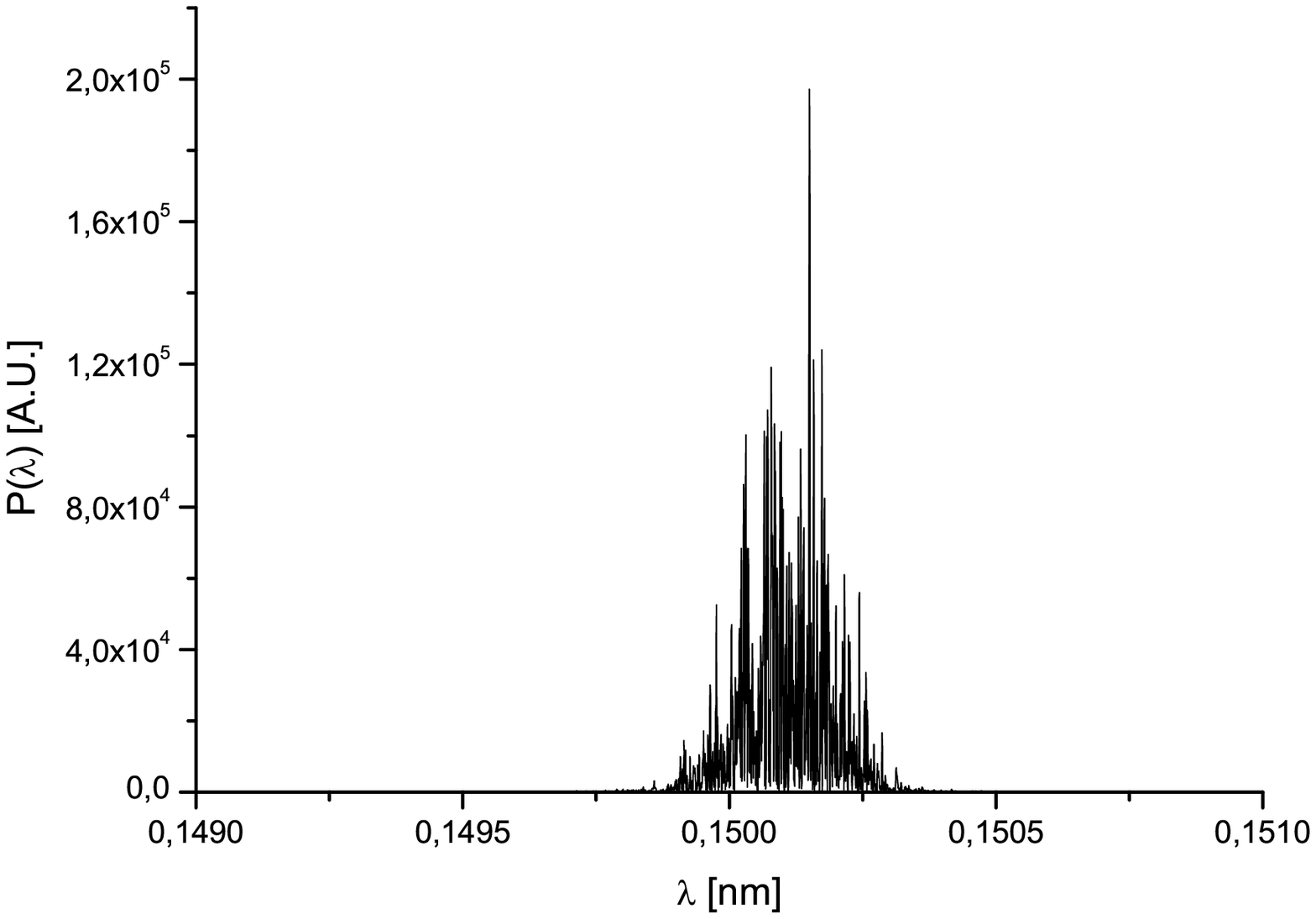}
\caption{Spectrum at the end of the fourth part of the undulator.
The first shutter is on. } \label{2spe}
\end{figure}

The power and spectrum after the fourth part of the undulator are
shown in Fig. \ref{2pow} and Fig. \ref{2spe}.

In this case the SASE process reaches saturation in the second
undulator part. The electron beam energy loss and induced energy
spread are severe, and prevent the beam to undergo SASE process
again in the last undulator part.
%

\subsection{Both shutters on}

\begin{figure}[tb]
\includegraphics[width=0.5\textwidth]{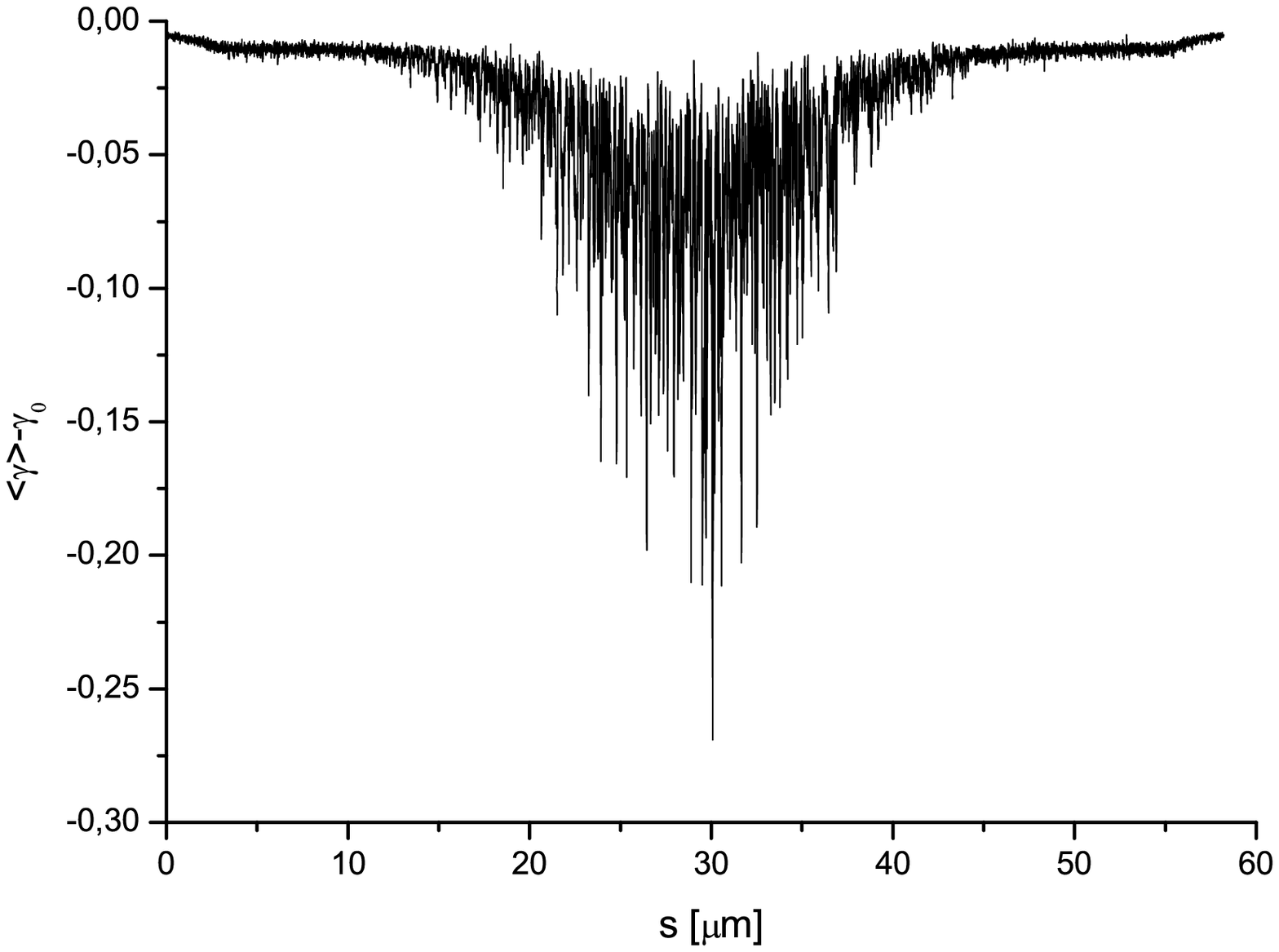}
\includegraphics[width=0.5\textwidth]{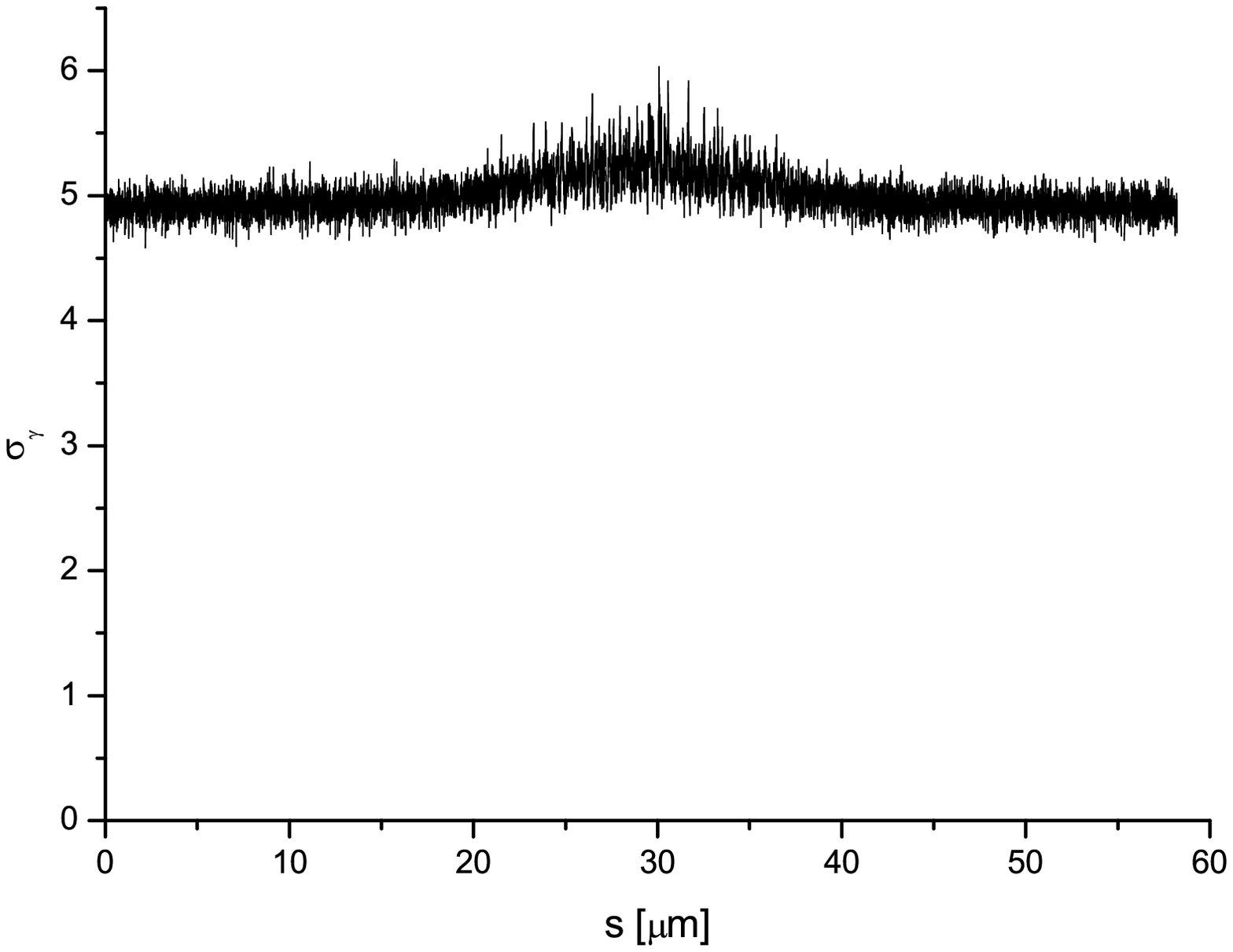}
\caption{Electron beam energy loss (left) and induced energy
spread (right) at the entrance of the fifth part of the undulator.
Both shutters are on.} \label{3enspr}
\end{figure}

Finally, we study the case when both shutters are on, Fig.
\ref{s2} bottom. In this case, the SASE process can start from
shot noise in the fifth part of the undulator (16 segments, $97.6$
m), at $0.1$ nm, because the presence of both shutters prevent
saturation before, and the beam quality is preserved up to the
entrance of the last undulator part, see Fig. \ref{3enspr}.

\begin{figure}[tb]
\includegraphics[width=1.0\textwidth]{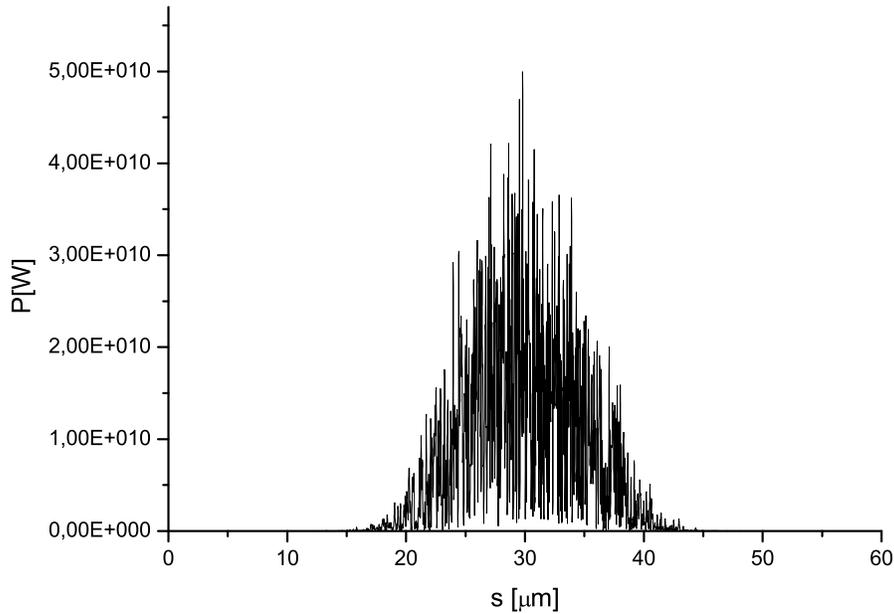}
\caption{Beam power distribution at the end of the fifth part of
the undulator. Both shutters are on. } \label{3pow}
\end{figure}

\begin{figure}[tb]
\includegraphics[width=1.0\textwidth]{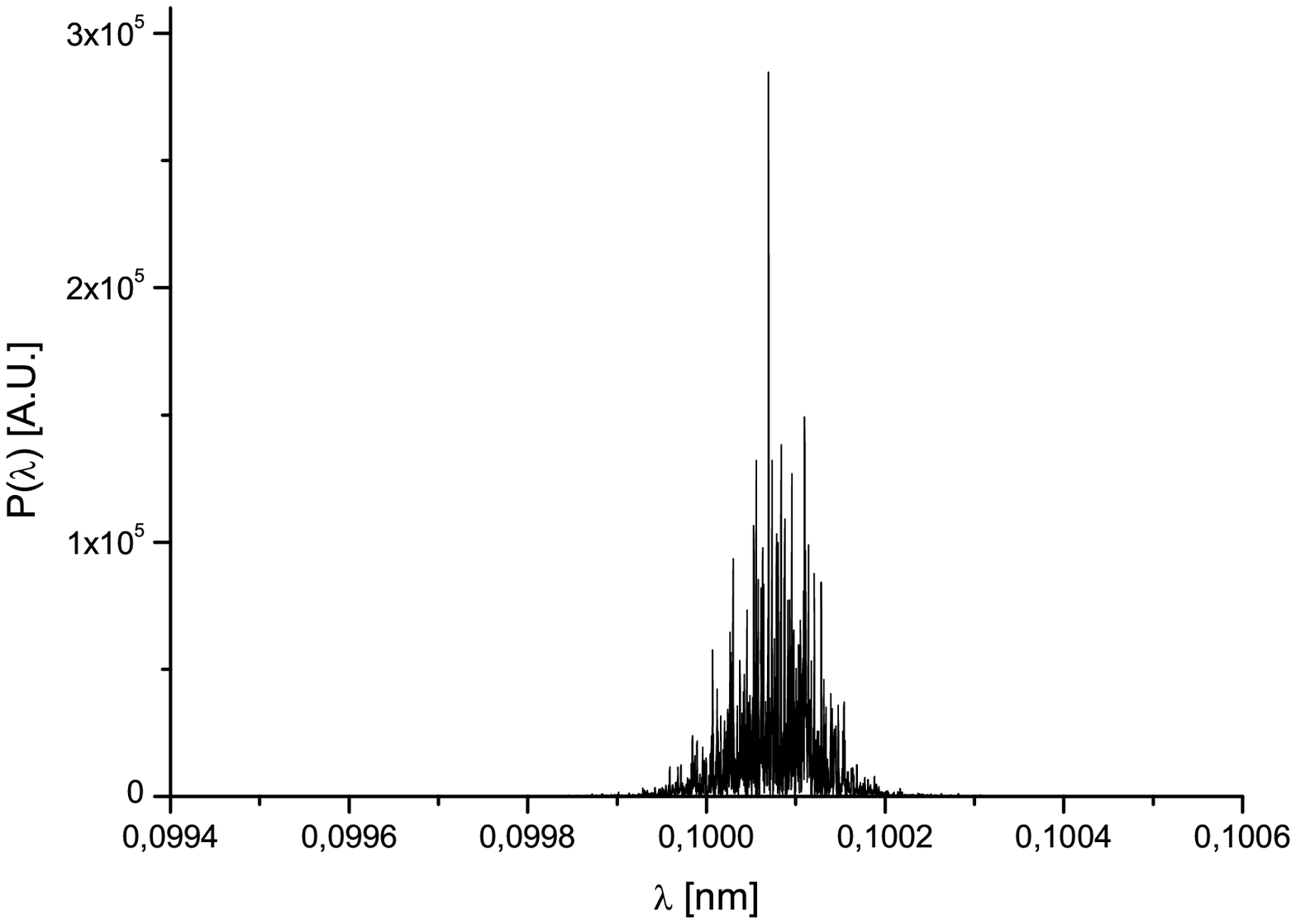}
\caption{Spectrum at the end of the fifth part of the undulator.
Both shutters are on.} \label{3spe}
\end{figure}

%
The power and spectrum after the fifth part of the undulator are
shown in Fig. \ref{3pow} and \ref{3spe}.

In this case the SASE process reaches saturation in the fifth
undulator part.

\section{Photon distribution}

\begin{figure}[tb]
\includegraphics[width=1.0\textwidth]{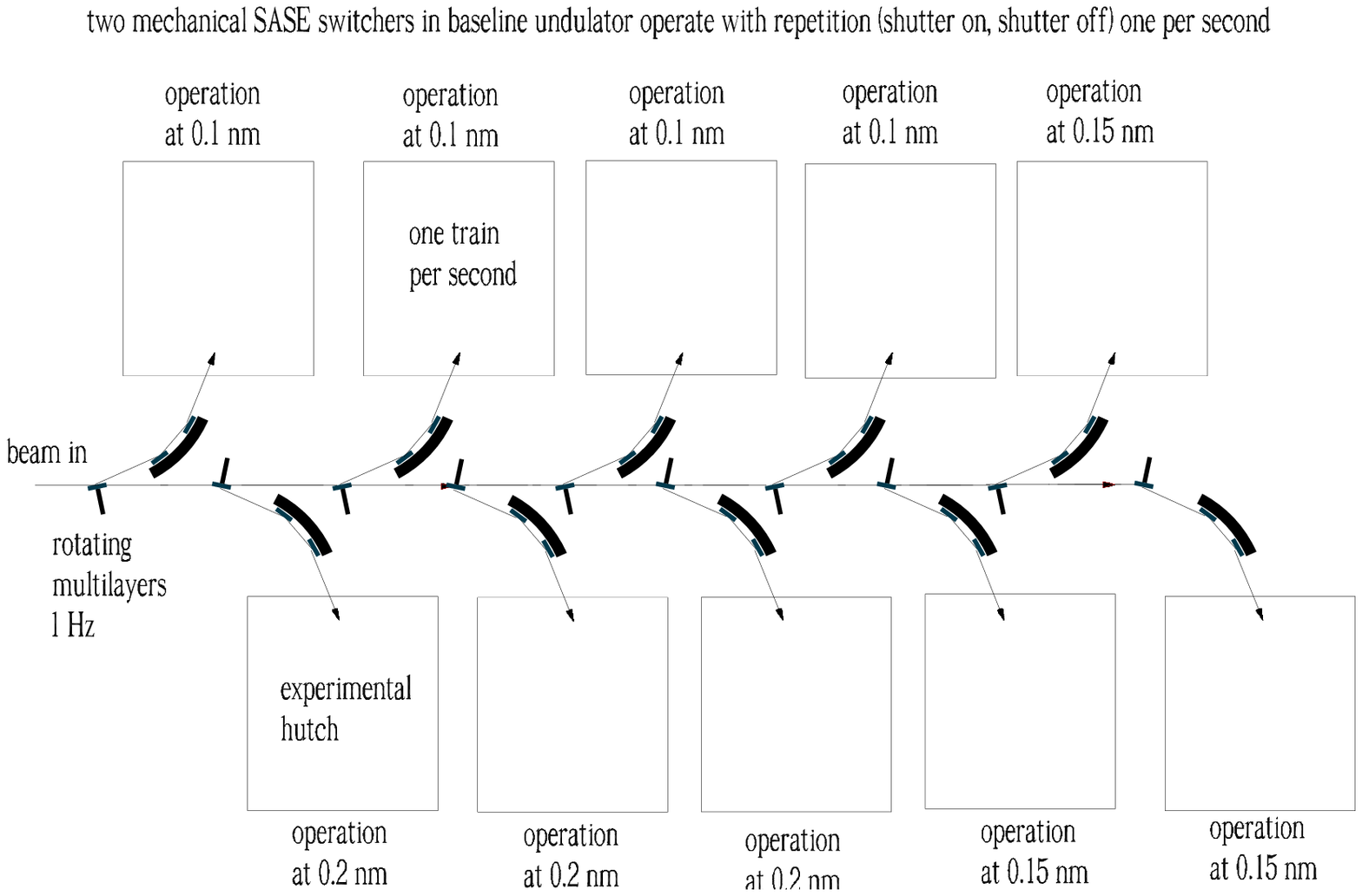}
\caption{Proposed SASE undulator beam line for multi-color mode
operation. A photon beam distribution system based on movable
multilayer X-ray mirrors. Distribution of photons is achieved on
the basis of pulse trains and it is possible to serve
simultaneously ten user stations with one train per second
repetition rate at three different wavelengths. In this case each
SASE shutter should be operated with one Hz repetition rate for a
single on-off-on cycle only.} \label{s5}
\end{figure}
As said before, the distribution of photons is done on the basis
of pulse-trains. The two mechanical shutters need to operate at a
frequency of $1$ Hz for a single on-off-on cycle with switching
time of less than $100$ ms (the temporal delay between two
consecutive trains). Consider a temporal interval of $1$ second,
i.e. $10$ trains of electron bunches. During the first $400$ ms,
the first three trains of electron bunches are let through with
both shutters off (see Fig. \ref{s2} upper part). Therefore, three
trains of radiation at $0.2$ nm are produced. Then, the first
shutter is switched on in less than $100$ ms, i.e. in the interval
between the next two trains. During the next $300$ ms, three
trains of electron bunches produce radiation at $0.15$ nm.
Finally, the second shutter is switched on during the interval
between the two following trains, and other four trains of
radiation are produces during the final $300$ ms, this time at
$0.1$ nm. Once separate color pulses are obtained in this way,
they can be distributed to different users. Combining this method
with a photon-beam distribution system based on movable multilayer
X-ray mirrors, as discussed before in \cite{OUR1}, can provide an
efficient way to generate a multi-user facility. This option,
exemplified in Fig. \ref{s5}, is not specific for the European
XFEL and may be applied for LCLS and other XFELs.

\begin{figure}[tb]
\includegraphics[width=1.0\textwidth]{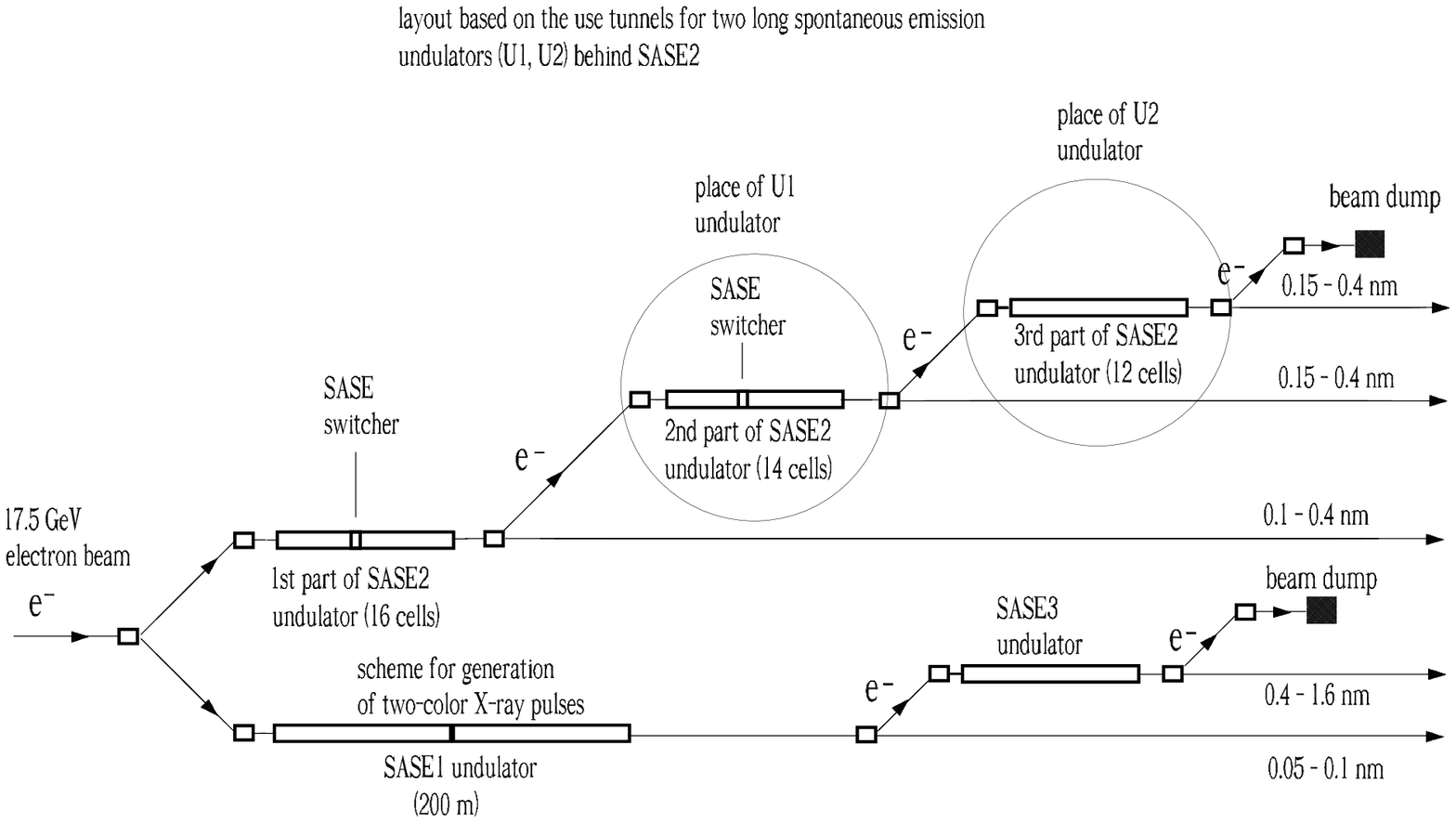}
\caption{Possible extension of the number of user stations which
can operate simultaneously at three different wavelengths at the
European XFEL. The present XFEL layout enables to accommodate two
long undulators behind SASE2, for spontaneous emission in
parasitic mode of operation. One may exploit these beamlines and
distribute the undulator modules of SASE2 respectively inside the
SASE2, U1, and U2 tunnels. Distribution of photons is achieved on
the basis of pulse trains. Two mechanical SASE switchers in the
first and second parts of the SASE2 undulator operate with
repetition frequency of $1$ Hz for a single on-off-on cycle.}
\label{s6}
\end{figure}
An option for the distribution of photons specific for the
European XFEL may also be considered. The original layout of the
European XFEL includes two long undulators for spontaneous
emission behind SASE2. These two undulators use the spent electron
beam of SASE2. We speculate on the possibility of distributing the
SASE2 undulator modules in three parts (14 cells, 12 cells and 16
cells), tuned at three different wavelengths as shown before,  and
of installing the second and the third part inside the U1 and U2
tunnels instead of the spontaneous emission undulators. The idea
is sketched in Fig. \ref{s6}. Combining this re-installation with
mechanical SASE switchers  for control of the FEL amplification
process can provide an efficient way to generate a multi-user
facility. The two mechanical shutters and the magnetic chicanes
would be installed as shown in Fig. \ref{s6}. Different colors may
then be transported to different experimental halls. In principle
then, at each experimental hall one may still take advantage of
the multi-user scheme in Fig. \ref{s5}.

\begin{figure}[tb]
\includegraphics[width=1.0\textwidth]{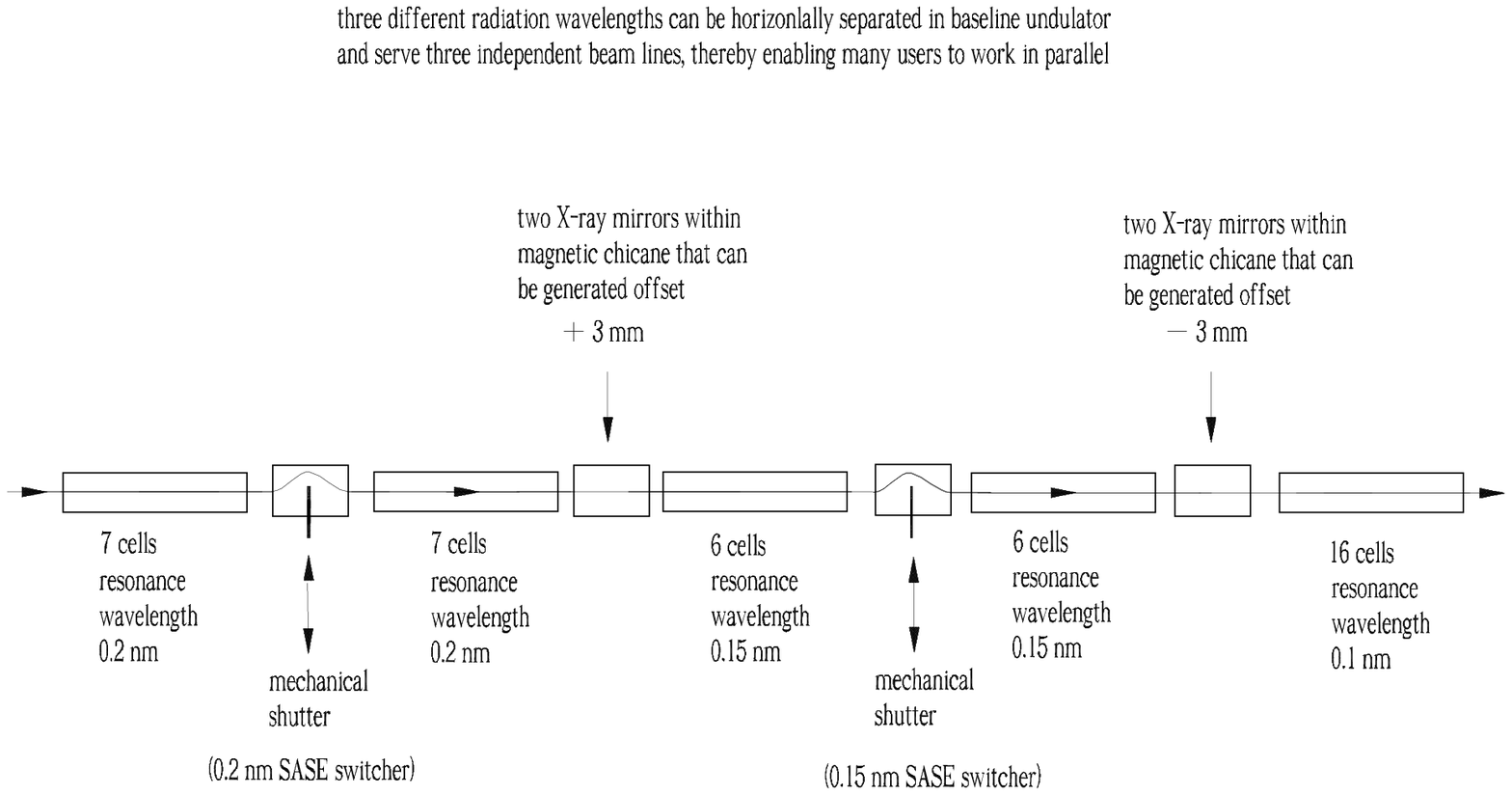}
\caption{Design of undulator system for generating in parallel
three color X-ray pulses. Three different radiation wavelengths
can be horizontally separated in baseline undulator and serve
three independent beam lines.} \label{s7}
\end{figure}

\begin{figure}[tb]
\includegraphics[width=1.0\textwidth]{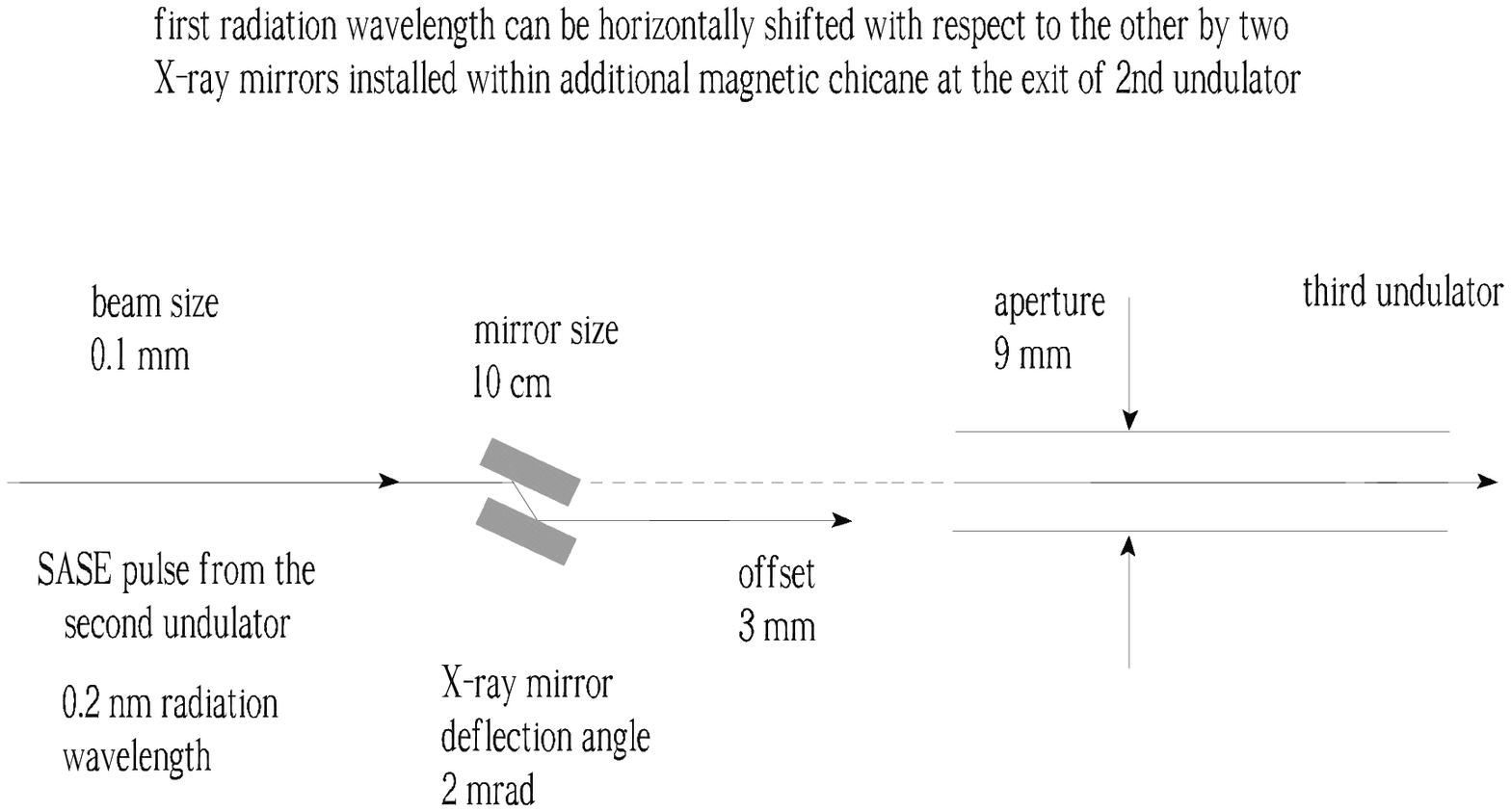}
\caption{Scheme for separating the first radiation wavelength with
respect to the other within undulator system. Two X-ray mirrors
can be installed within  additional magnetic chicane at the second
undulator exit.} \label{s8}
\end{figure}

\begin{figure}[tb]
\includegraphics[width=1.0\textwidth]{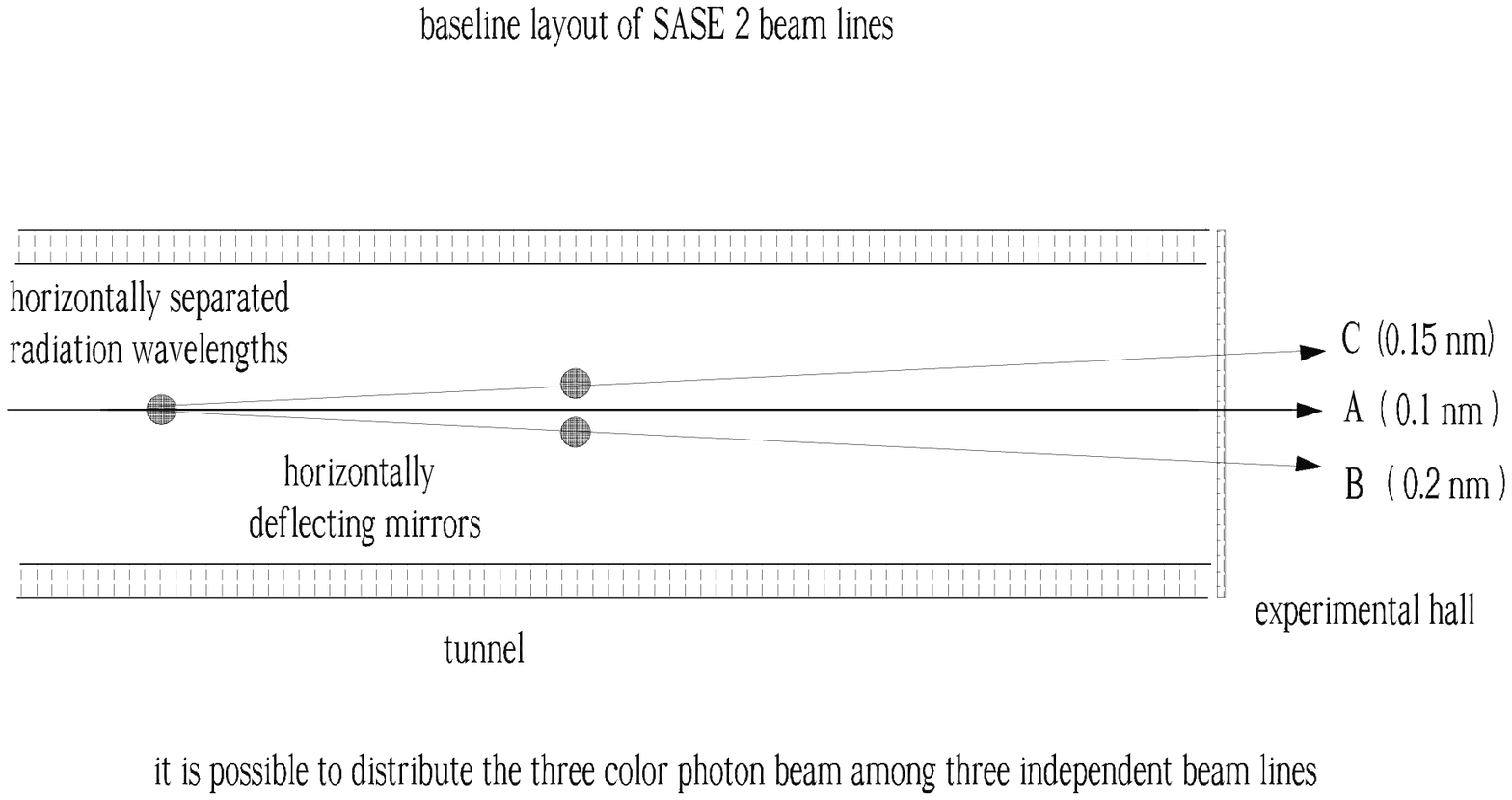}
\caption{Baseline layout of the SASE 2 beam lines. It is possible
to distribute the three-color photon beam among three independent
beam lines. Distribution of photons is achieved on the basis of
the horizontal separation of different colors in the baseline
undulator.} \label{s9}
\end{figure}
Finally, similarly as in \cite{OUR0} we remark that the main
difficulty concerning the distribution of photons consists in the
separation of the three colors. Once this task is performed,
mirrors in the photon beam transport system can be used to
distribute the three-color photon beam among three independent
user beam lines. As in \cite{OUR0} we propose to separate the
three colors already in the undulator with the help of x-ray
mirrors. The idea is sketched in Fig. \ref{s7}, Fig. \ref{s8} and
Fig. \ref{s9}. The three colors can be separated horizontally in
two stages by installing the two-mirror setup sketched in Fig.
\ref{s8} after the undulator parts producing a given color, as
specified in Fig. \ref{s7}. Note that the installation of these
setups also requires the presence of a weak magnetic chicane in
order to create an offset in the electron trajectory to accomodate
the mirrors. The horizontal offset should be chosen small enough
to account for the presence of the spontaneous radiation absorbers
in the vacuum chamber, which limits the effective aperture to a
circle of $9$ mm diameter. The horizontal offset may be therefore
chosen to be around $3$ mm, which is enough for separating the
two-color pulses: in fact, at the position of the optical station
the FWHM beam size is less than a millimeter. Additionally,
mirrors can also be used to generate a few $\mu$rad
deflection-angle, which is not important within the undulator but
will create further a small extra-separation of a few millimeters
at the position of the experimental station, as shown in Fig.
\ref{s9}.

\section{Conclusions}

We presented a novel method to control the SASE amplification
process in the baseline XFEL undulator with the help of mechanical
SASE switchers. After the lasing of LCLS \cite{LCLS2}, a new
scenario where the beam formation system works as in the ideal
case has become reality. This allows for a reduction of the gain
length in the SASE process and for exploitation of the
extra-available undulator modules. In particular, we show how it
is possible to accommodate three FELs, lasing at three different
wavelengths within the foreseen undulator length for the SASE2
beam line at the European XFEL. The scheme makes use of mechanical
switchers capable of switching on and off the SASE process at a
given particular wavelength. Three possible configurations of two
switchers allows for separate production of each of the three
wavelengths. The switchers should work on the basis of a train of
pulses, with a frequency of $1$ Hz for an on-off-on cycle, and
with a switching time of less than $100$ ms. In this way,
simultaneous operation at three different wavelength is possible.
Distribution of the photons to different stations is discussed.

\section{Acknowledgements}

We are grateful to Massimo Altarelli, Reinhard Brinkmann, Serguei
Molodtsov and Edgar Weckert for their support and their interest
during the compilation of this work.

\end{document}